\documentclass[prl,twocolumn,floatfix,showpacs ]{revtex4-1}
\UseRawInputEncoding
\usepackage{amsmath}
\usepackage{amssymb}
\usepackage{graphicx}
\usepackage{dcolumn}
\usepackage{bm}
\usepackage[colorlinks=true,linkcolor=blue,anchorcolor=blue, citecolor=cyan,urlcolor=cyan]{hyperref}
\usepackage[mathlines]{lineno}
\usepackage{ulem}
\usepackage{epstopdf}
\begin{document}
\title{Zero-net-magnetization hybrid magnet}
\author{San-Dong Guo}
\email{sandongyuwang@163.com}
\affiliation{School of Electronic Engineering, Xi'an University of Posts and Telecommunications, Xi'an 710121, China}
\author{Shi-Hao Zhang}
\email{zhangshh@hnu.edu.cn}
\affiliation{School of Physics and Electronics, Hunan University, Changsha 410082, China}
\begin{abstract}
Zero-net-magnetization magnets possess ultradense and ultrafast application potential, benefiting from their intrinsic zero stray field and terahertz dynamics characteristics.  Herein, we propose the concept of  zero-net-magnetization hybrid magnet, in which magnetic atoms with opposite spin polarization are partially coupled via spatial inversion ($P$) symmetry, partially via  rotation/mirror ($C/M$) symmetry or  partially without any symmetry correlation.
From a local perspective and neglecting the interactions between local regions, hybrid magnet can be regarded as being composed of $PT$-antiferromagnet  (possessing the combined symmetry ($PT$) of $P$ and time-reversal ($T$)), altermagnet, or fully compensated ferrimagnet. To realize hybrid magnet, we propose that such system can be constructed by forming heterojunction with three types of zero-net-magnetization magnetic monolayers. We mainly investigate the heterojunction  composed of two kinds of zero-net-magnetization magnets, among which one type corresponds to fully compensated ferrimagnet. When heterojunction  hybrid magnet exhibits a type-II band alignment, only one of electron doping and hole doping can induce a net magnetic moment, while the other hardly generates any net magnetization. Taking the heterojunction constructed by $PT$-antiferromagnet and fully compensated ferrimagnet as an example, we verify our proposal by means of the tight-binding (TB) model.
Finally, taking the   $\mathrm{Cr_2C_2S_6}$/$\mathrm{CrMoC_2S_6}$  heterojunction as an example, we perform first-principles calculations combined with electric field modulation to validate our TB model and theoretical proposal.

\end{abstract}
\maketitle
\textcolor[rgb]{0.00,0.00,1.00}{\textbf{Introduction.---}}
Zero-net-magnetization magnets possess nearly zero macroscopic magnetic moment and extremely weak stray magnetic fields. Their prominent advantages include minimal magnetic interference, excellent immunity to external magnetic disturbances, low energy consumption, and ultrafast spin dynamics. Such magnets are highly suitable for high-density integration, high-speed spintronic devices and precision electronic systems\cite{k1,k2,k3}. The main types cover  $PT$-antiferromagnet  (possessing the combined symmetry ($PT$) of  spatial inversion  ($P$) and time-reversal ($T$)), altermagnet, and  fully compensated ferrimagnet, all of which realize net magnetization cancellation through special lattice symmetry or appropriate electron filling\cite{k4,k5,k6,k7,k7-1}.
$PT$-antiferromagnets represent a conventional class of zero-net-magnetization magnet. They exhibit global spin degeneracy throughout the entire Brillouin zone (BZ), which inherently suppresses the magneto-optical response, anomalous Hall effect, and anomalous valley Hall effect.

Altermagnets have recently emerged as a  third fundamental magnetic category beyond conventional ferromagnets and antiferromagnets, which are  characterized by collinear alternating spin arrangements and special lattice symmetry  (rotation/mirror ($C/M$) symmetry) compensation\cite{k4,k5}. An altermagnet can be described by the spin group $[C_2\parallel O]$\cite{lqh1}, where $C_2$ denotes a twofold rotation in spin space (equivalent to spin inversion) with its axis perpendicular to the collinear spin direction, while $O$  represents the symmetry operations in real space. Distinct from conventional
$PT$-antiferromagnets, however, their intrinsic symmetry breaking enables prominent spin polarization, nontrivial band splitting, and abundant emergent transport properties, such as the anomalous Hall effect and magneto-optical responses\cite{k8}. Numerous altermagnets have been theoretically predicted and experimentally synthesized, greatly advancing the rapid development of this research field\cite{k4,k5,k8,k6-0,k6-1,k6-2,k6-3,k6-311, ex0,ex01,ex02,ex1,ex2,ex3,ex4,ex51,ex52}. Recently, hidden altermagnetism has been formally proposed and subsequently verified by experimental measurements\cite{h6,ex5,ex53}.

Fully compensated ferrimagnets constitute another category of zero-net-magnetization magnet. Unlike other magnetic systems, their zero-net-magnetization is not guaranteed by lattice symmetry, but instead achieved through appropriate electron filling\cite{k6,k7,k7-1}.
Like some altermagnets, fully compensated ferrimagnets can also host a variety of intriguing physical phenomena, such as the anomalous Hall effect, anomalous Nernst effect, nonrelativistic spin-polarized currents, and the magneto-optical Kerr effect. Fully compensated ferrimagnets can be achieved by breaking the lattice symmetry that links the two spin sublattices in $PT$-antiferromagnets and altermagnets. Typical strategies include applying an electric field, alloying construction, heterostructure design, and spin-order modulation\cite{k6,k7,qq3,qq4,qq5}. Finally, it is worth pointing out that, from the perspective of symmetry, fully compensated ferrimagnets should be classified into the ferromagnetic (FM) family\cite{lqh}.

\begin{figure}[t]
    \centering
    \includegraphics[width=0.48\textwidth]{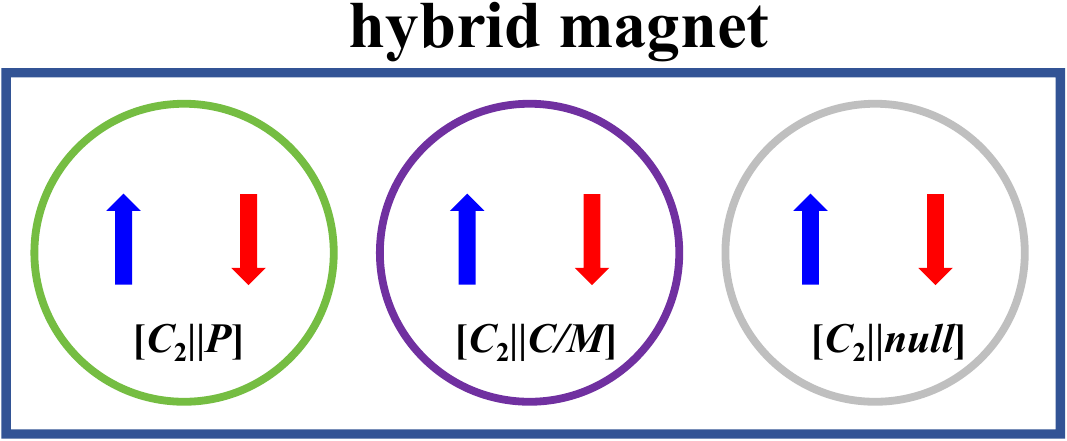}
    \caption{(Color online) the schematic of the zero-net-magnetization hybrid magnet, where the system is divided into several sectors. For each sector, the  magnetic atoms with opposite spin polarization are  connected via $P$ symmetry $[C_2\parallel P]$,  via $C/M$ symmetry  $[C_2\parallel C/M]$, or  without any symmetry connection  $[C_2\parallel null]$. }\label{a0}
   \end{figure}

It is therefore natural to ask whether such an electronic state can exist when a material is divided into multiple sectors, with different sector hosting
$PT$-antiferromagnet, altermagnet, or fully compensated ferrimagnet, respectively.  When the  interaction among different sectors is sufficiently weak, such electronic states can indeed exist. Accordingly, we herein propose the concept of zero-net-magnetization hybrid magnet. Based on the tight-binding (TB) model and first-principles calculations, the proposed scheme and its related physical effects are systematically validated. This study broadens the scope of zero-net-magnetization magnets and facilitates the in-depth investigation of associated physical effects.

\begin{figure}[t]
    \centering
    \includegraphics[width=0.48\textwidth]{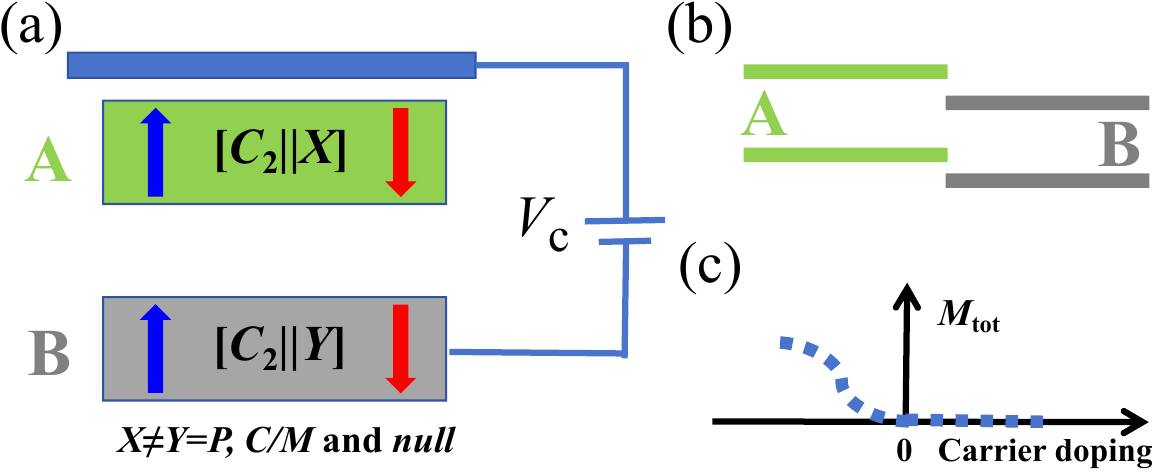}
    \caption{(Color online) (a): two zero-net-magnetization magnets  (A and B) are stacked via vdW interaction to form a heterojunction, and their magnetic atoms with opposite spins are connected through  [$C_2$$\parallel$$X/Y$] ($X$$\neq$$Y$=$P$, $C/M$ or $null$)  symmetry. (b):the schematic diagram of a type-II heterojunction. The energy bands of material A and material B are staggered, leading to spatial separation of electrons and holes. (c): when the carrier type is tuned by gate voltage $V_c$ (a), if A is an FC-FIM semiconductor and B is a $PT$-AFM  semiconductor, moderate electron doping cannot generate a net magnetic moment, while hole doping will induce the formation of a net magnetic moment. }\label{a}
   \end{figure}

\textcolor[rgb]{0.00,0.00,1.00}{\textbf{Concept of hybrid magnet.---}}
A zero-net-magnetization hybrid magnet refers to a material containing magnetic atoms with opposite spin polarization, which are partially connected via $P$ symmetry, partially via $C/M$ symmetry, or  partially without any symmetry connection (the material can be divided into several sectors.), as shown in \autoref{a0}. When ignoring the interactions among sectors, the material can be regarded as composed of $PT$-antiferromagnets, altermagnets, or fully compensated ferrimagnets. In reality, however, the presence of intrinsic interactions makes it difficult to well define these zero-net-magnetization magnets. To achieve their approximate definition and characterization, layered materials serve as promising candidates.
Here, we only consider the mixing of two types of zero-net-magnetization magnets, which can be classified into three cases: the mixture of $PT$-antiferromagnets and altermagnets, the mixture of $PT$-antiferromagnets and fully compensated ferrimagnets, and the mixture of altermagnets and fully compensated ferrimagnets.
From the perspective of artificial stacking design, heterojunctions provide a feasible strategy (see \autoref{a} (a)), and  there is a weak van der Waals (vdW) interaction between the two layers.

Heterojunctions can be classified into three types according to their band alignment characteristics\cite{hz}: type-I heterojunction (straddling alignment),
type-II heterojunction (staggered alignment) and type-III heterojunction (broken-gap alignment). Among them, the energy bands of  type-II heterojunctions near the Fermi level  can fully involve both constituent materials. The energy bands of material A and material B are staggered (see \autoref{a} (b)), which results in the spatial separation of electrons and holes that accumulate on either side of the two materials. From the perspective of carrier doping, $PT$-antiferromagnets and altermagnets exhibit no net magnetic moment, while fully compensated ferrimagnets can generate a net magnetic moment. For $PT$-antiferromagnets and altermagnets, the spin-up and spin-down energy bands exhibit the following relationship:
\begin{equation}\label{d-1}
E_{\uparrow}(\vec{k})=[C_2\parallel O]E_{\uparrow}(\vec{k})=E_{\downarrow}(O\vec{k})
\end{equation}
where $O$ denotes $P$ or  $C/M$ symmetry.  And then, we obtain:
\begin{equation}\label{d-2}
g_{\uparrow}(E)=\sum_{\vec{k}}\delta[E-E_{\uparrow}(\vec{k})]=\sum_{\vec{k}}\delta[E-E_{\downarrow}(O\vec{k})]
\end{equation}
When $\vec{k}$ spans the entire first BZ, $O\vec{k}$ also spans the entire BZ. Therefore, \autoref{d-2} can be rewritten as:
\begin{equation}\label{d-3}
\sum_{\vec{k}}\delta[E-E_{\downarrow}(O\vec{k})]=\sum_{\vec{k}^{\prime}}\delta[E-E_{\downarrow}(\vec{k}^{\prime})]=g_{\downarrow}(E)
\end{equation}
where $g_{\uparrow}(E)$ and  $g_{\downarrow}(E)$  denote spin-up and spin-down density of states (DOS).  According to \autoref{d-2} and \autoref{d-3}, $g_{\uparrow}(E)$ and  $g_{\downarrow}(E)$  are identical ($g_{\uparrow}(E)$=$g_{\downarrow}(E)$) in  $PT$-antiferromagnets and altermagnets, but they are generally not equal ($g_{\uparrow}(E)$$\neq$$g_{\downarrow}(E)$)  in fully compensated ferrimagnets.

In general, the net magnetization  $M_{tot}$ in the magnet can be written as:
\begin{equation}\label{d-4}
M_{tot}=\mu_B\int_{-\infty}^{E_F}(g_{\uparrow}(E)-g_{\downarrow}(E))dE
\end{equation}
where $\mu_B$ is the Bohr magneton, $E_F$ is the Fermi level.
For $PT$-antiferromagnets,  altermagnets and fully-compensated ferrimagnets,  $M_{tot}$ is equal to zero. Nevertheless, when shifting the Fermi level $E_F$ to vary the charge-carrier concentration, $M_{tot}$ remains exactly zero in $PT$-antiferromagnets and altermagnets  due to $g_{\uparrow}(E)$=$g_{\downarrow}(E)$, whereas it is generally non-zero in fully compensated ferrimagnets  due to $g_{\uparrow}(E)$$\neq$$g_{\downarrow}(E)$.

For a type-II heterojunction with one layer being a fully compensated ferrimagnet, gate voltage can be employed to modulate the concentration and type of charge carriers, and then, for example, the  hole doping can give rise to a net magnetic moment, while no net magnetic moment is formed under electron doping  (see \autoref{a} (c)). To realize such effect of magnetic moment modulation via carrier type, a feasible implementation scheme is to construct a heterojunction composed of an fully compensated ferrimagnetic (FC-FIM)  monolayer and a $PT$-antiferromagnetic (AFM) monolayer. Meanwhile, the type-II band alignment of the heterojunction can be achieved by out-of-plane electric field modulation, even if the heterojunction possesses an intrinsic type-I or type-III band alignment. In what follows, we elaborate on this mechanism with a concrete  TB model.

\begin{figure}[t]
    \includegraphics[width=0.45\textwidth]{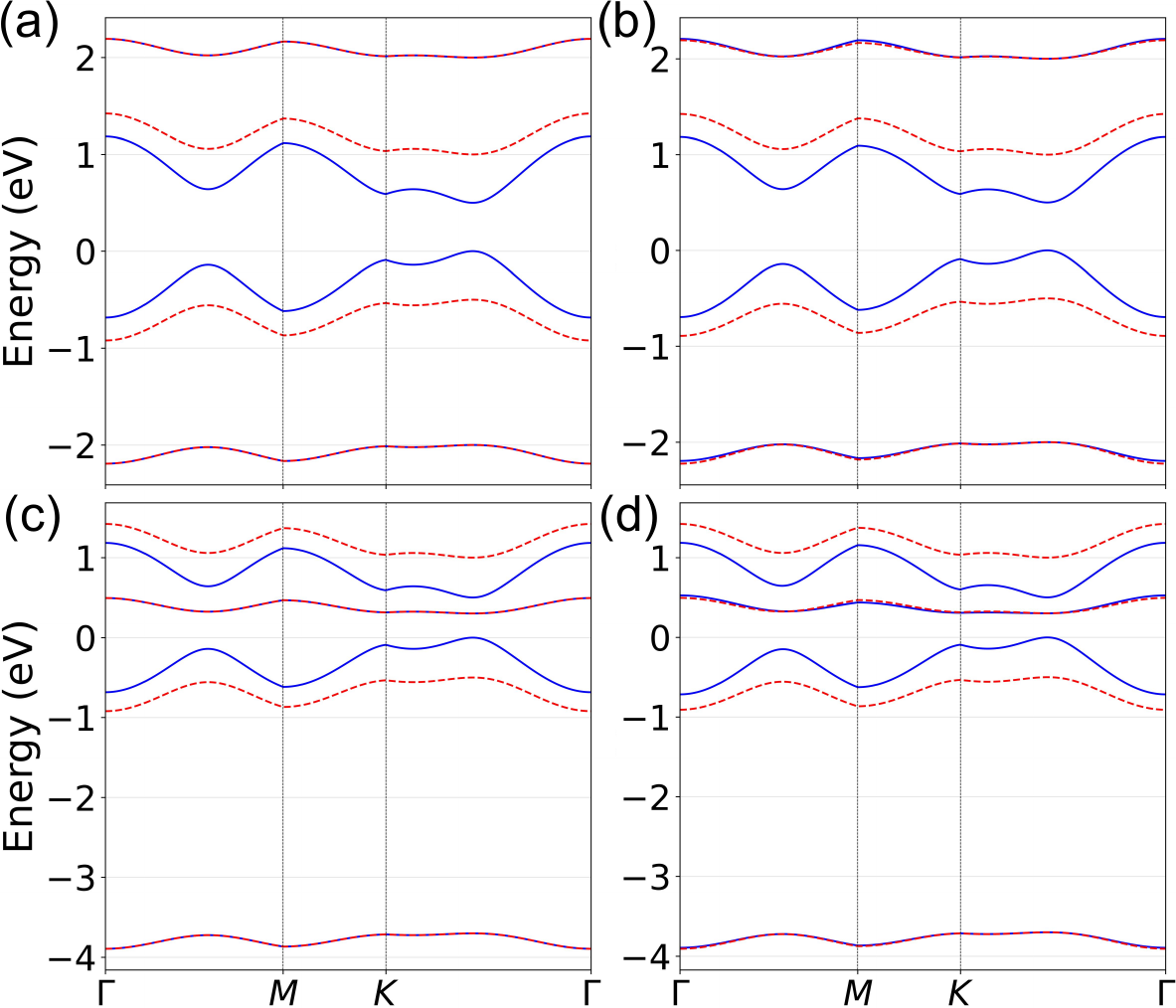}
    \caption{(Color online) The energy bands obtained from minimal effective Hamiltonian. We set $\varepsilon_1=$0.0\,eV, $\varepsilon_2=$0.5\,eV, $\Delta_1=$2.0\,eV, $\Delta_2=$0.5\,eV, $t_1=t_2=$0.3\,eV. (a) $\Delta_U=$0.0, and $t_\perp=t_\perp'=$0.0. (b) $\Delta_U=$0.0, and $t_\perp=$0.05\,eV, $t_\perp'=$0.01\,eV. (c) $\Delta_U=$-1.7\,eV, and $t_\perp=t_\perp'=$0.0. (d) $\Delta_U=$-1.7\,eV, and $t_\perp=$0.05\,eV, $t_\perp'=$0.01\,eV. The blue solid lines and red dotted lines represent spin-up and spin-down states, respectively.}
    \label{a1}
\end{figure}
\begin{figure*}[t]
    \centering
    \includegraphics[width=0.96\textwidth]{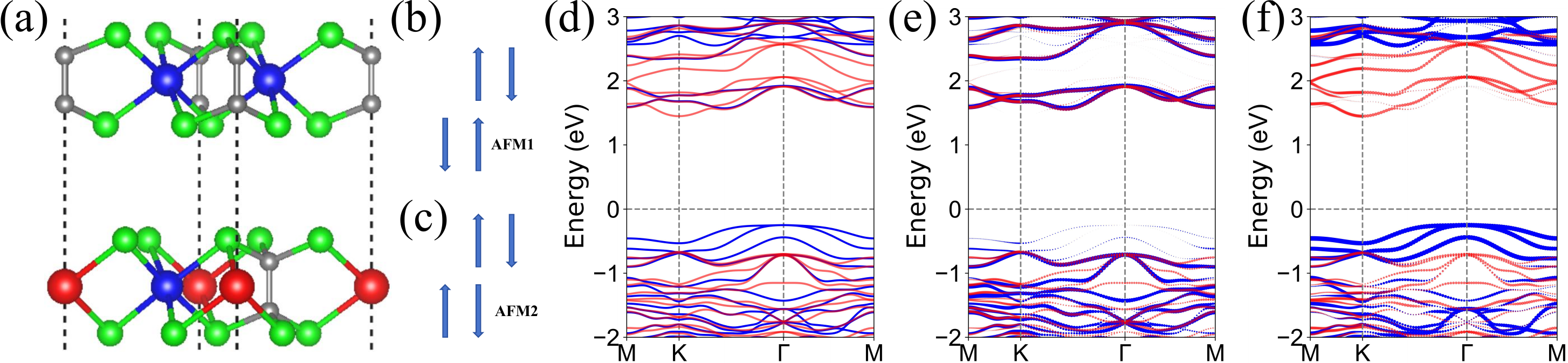}
    \caption{(Color online) For  $\mathrm{Cr_2C_2S_6}$/$\mathrm{CrMoC_2S_6}$ heterojunction,
(a): the crystal structure; (b,c): two possible magnetic configurations AFM1 and AFM2;
(d): the global spin-polarized band structure;
(e,f): the spin-polarized band structure projected onto $\mathrm{Cr_2C_2S_6}$ and $\mathrm{CrMoC_2S_6}$. In (a),  the blue,  red, green and gray balls  represent Cr, Mo, S and C atoms, respectively. In  (d, e, f), the spin-up
and spin-down channels are depicted in blue and red.  }\label{b}
\end{figure*}

\textcolor[rgb]{0.00,0.00,1.00}{\textbf{Effective Hamiltonian model.---}}
To construct a minimal effective  TB model, we study a Bernal-stacked hexagonal bilayer lattice with lattice constant $a$, composed of an FC-FIM layer and a $PT$-AFM layer. The unit cell hosts four orbitals. The first layer contains two equivalent magnetic atoms, $\{1A, 1B\}$  (\{Cr, Cr\}), arranged in an AFM sublattice pattern (opposite spin orientations). The second layer comprises two inequivalent atoms, $\{2A, 2B\}$  (\{Cr, Mo\}), also forming an AFM sublattice (opposite spins). In this heterostructure, the $1A$ site is positioned directly above $2A$, while $1B$ lies above a hollow site of the second layer. Both layers are magnetic insulators with a finite bulk band gap. The system realizes a type-I band alignment under no external electric field, where both the valence band maximum (VBM) and conduction band minimum (CBM) originate from the FC-FIM layer.

The effective Hamiltonian can be constructed within the basis $\Psi_{\boldsymbol{k}\sigma} = [c_{1A\sigma}, c_{1B\sigma}, c_{2A\sigma}, c_{2B\sigma}]^T$. Here $\sigma$ refers to spin index. Then the $4\times4$ Hamiltonian $H_\sigma(\boldsymbol{k})$ can be written as
{\small
\begin{equation}
H_\sigma(\boldsymbol{k}) =
\begin{pmatrix}
\varepsilon_1 + \sigma\Delta_1 & -t_1 f & -t_\perp f & -t_\perp f^*\\
-t_1 f^* & \varepsilon_1 - \sigma\Delta_1 & 0 & -t_\perp' f\\
-t_\perp f & 0 & \varepsilon_1 + \sigma\Delta_2 + \Delta_U & -t_2 f\\
-t_\perp f & -t_\perp' f^* & -t_2 f^* & \varepsilon_2 - \sigma\Delta_2 + \Delta_U
\end{pmatrix}
\end{equation}
}
where $\varepsilon_1$ and $\varepsilon_2$ denote the on-site energies of the Cr ($\{1A, 1B, 2A\}$) and Mo ($\{2B\}$) atoms, respectively. $\Delta_1$ and $\Delta_2$ represent the AFM spin splittings in the two layers. $\Delta_U$ is the potential difference between the two layers generated by an external electric field. $t_1$ and $t_2$ describe intralayer hoppings, while $t_\perp$ and $t_\perp'$ correspond to interlayer hoppings. The hexagonal nearest-neighbor structure factor in the hopping tern is:
\begin{equation}
f(\boldsymbol{k}) = 2e^{ik_x a/2}\cos\left(\frac{\sqrt{3}k_y a}{2}\right) + e^{-ik_x a}.
\end{equation}

At the K point, the layers are fully decoupled because the structure factor $f(\boldsymbol{K})=0$. The band edges are:
\begin{align}
    {\rm VBM}_1 &= \varepsilon_1 - \Delta_1, \quad {\rm CBM}_1 = \varepsilon_1 + \Delta_1, \\
    {\rm VBM}_2 &= \min\left(\varepsilon_1-\Delta_2+\Delta_U,\ \varepsilon_2-\Delta_2+\Delta_U\right), \\
    {\rm CBM}_2 &= \max\left(\varepsilon_1+\Delta_2+\Delta_U,\ \varepsilon_2+\Delta_2+\Delta_U\right).
\end{align}
The critical $\Delta_U$ values for the transition of type-I into type-II transition are:
\begin{align}
    \Delta_{U{\rm c1}}^{(0)} &= \varepsilon_1 + \Delta_1 - \max\left(\varepsilon_1+\Delta_2,\ \varepsilon_2+\Delta_2\right), \label{eq:dc1_0} \\
    \Delta_{U{\rm c2}}^{(0)} &= \varepsilon_1 - \Delta_1 - \min\left(\varepsilon_1-\Delta_2,\ \varepsilon_2-\Delta_2\right).
\end{align}
The type-I alignment exists for $\Delta_{U{\rm c2}}^{(0)} < \Delta_U < \Delta_{U{\rm c1}}^{(0)}$. When out-of-plane electric field induce $\Delta_U > \Delta_{U{\rm c1}}^{(0)}$ or $\Delta_U < \Delta_{U{\rm c2}}^{(0)}$, the system possesses the transition from type-I alignment to type-II alignment.

At the $\Gamma$ point, the interlayer coupling $|f(\boldsymbol{\Gamma})| = 3$ reaches maximum interlayer coupling. Then we utilize the second-order energy correction for nondegenerate states:
\begin{equation}
    \Delta E_n^{(2)} = \sum_{m\neq n} \frac{\left|\langle n | H_\perp | m \rangle\right|^2}{E_n^{(0)} - E_m^{(0)}}.
\end{equation}
We obtain the corrected critical potential differences at $\Gamma$ point:
\begin{align}
    \Delta_{U{\rm c1}} &= \Delta_{U{\rm c1}}^{(0)} - \delta U_{\rm c1}, \label{eq:dc1_full} \\
    \Delta_{U{\rm c2}} &= \Delta_{U{\rm c2}}^{(0)} + \delta U_{\rm c2}, \label{eq:dc2_full}
\end{align}
where the perturbation corrections are:
\begin{align}
    \delta U_{\rm c1} &= 9\max\left(
    \frac{t_\perp^2}{E_{2A}^{(0)} - E_{1A}^{(0)}},\
    \frac{t_\perp^2}{E_{2B}^{(0)} - E_{1A}^{(0)}} + \frac{t_\perp'^2}{E_{2B}^{(0)} - E_{1B}^{(0)}}
    \right), \label{eq:du_c1} \\
    \delta U_{\rm c2} &= 9\max\left(
    \frac{t_\perp^2}{E_{1A}^{(0)} - E_{2A}^{(0)}},\
    \frac{t_\perp^2}{E_{1A}^{(0)} - E_{2B}^{(0)}} + \frac{t_\perp'^2}{E_{1B}^{(0)} - E_{2B}^{(0)}}
    \right). \label{eq:du_c2}
\end{align}
Here the unperturbed on-site energies are:
\begin{align*}
    E_{1A}^{(0)} &= \varepsilon_1+\Delta_1, \quad E_{1B}^{(0)} = \varepsilon_1-\Delta_1, \\
    E_{2A}^{(0)} &= \varepsilon_1+\Delta_2+\Delta_U, \quad E_{2B}^{(0)} = \varepsilon_2-\Delta_2+\Delta_U.
\end{align*}

Now we calculate the energy bands based on our effective Hamiltonian. The calculated band structures are shown in the \autoref{a1}. Four cases are considered, including the absence of an electric field and interlayer interaction (\autoref{a1} (a)), the absence of an electric field but the presence of interlayer interaction (\autoref{a1} (b)), the presence of an electric field but the absence of interlayer interaction (\autoref{a1} (c)), and the presence of both an electric field and interlayer interaction (\autoref{a1} (d)). It is clearly demonstrated that an external electric field can effectively drive the transition of the heterojunction from type-I to type-II (\autoref{a1} (a) to \autoref{a1} (c), or \autoref{a1} (b) to \autoref{a1} (d)). For weak interlayer interaction, the type-II heterojunction exhibits nearly spin degeneracy of the conduction bands near the Fermi level, while obvious spin splitting occurs in the valence bands (see \autoref{a1} (d)).

\textcolor[rgb]{0.00,0.00,1.00}{\textbf{Computational detail.---}}
The spin-polarized density functional theory\cite{1} calculations are  performed
using the Vienna ab initio simulation package (VASP)\cite{pv1,pv2,pv3} by using the projector augmented-wave (PAW) method. We use the generalized gradient approximation (GGA) proposed by
Perdew, Burke, and Ernzerhof (PBE)\cite{pbe} as  the exchange-correlation functional. The  kinetic energy cutoff of 500 eV,  total energy  convergence criterion of  $10^{-8}$ eV and  force convergence criterion of 0.001 $\mathrm{eV\cdot{\AA}^{-1}}$ are set to obtain reliable results.
We add Hubbard correction with $U$=3.00 eV\cite{qq3,f7-1} for $d$-orbitals of Cr and Mo atoms within the
rotationally invariant approach proposed by Dudarev et al\cite{du}.
A vacuum slab larger than 16 $\mathrm{{\AA}}$ is introduced to eliminate interlayer physical interactions between periodic supercells.
A 13$\times$13$\times$1 Monkhorst-Pack $k$-point meshes are adopted to sample the Brillouin zone  for structural relaxation and electronic structure calculations. The dispersion-corrected DFT-D3 method\cite{dft3} is employed to describe the vdW interactions.

\begin{figure*}[t]
    \centering
    \includegraphics[width=0.96\textwidth]{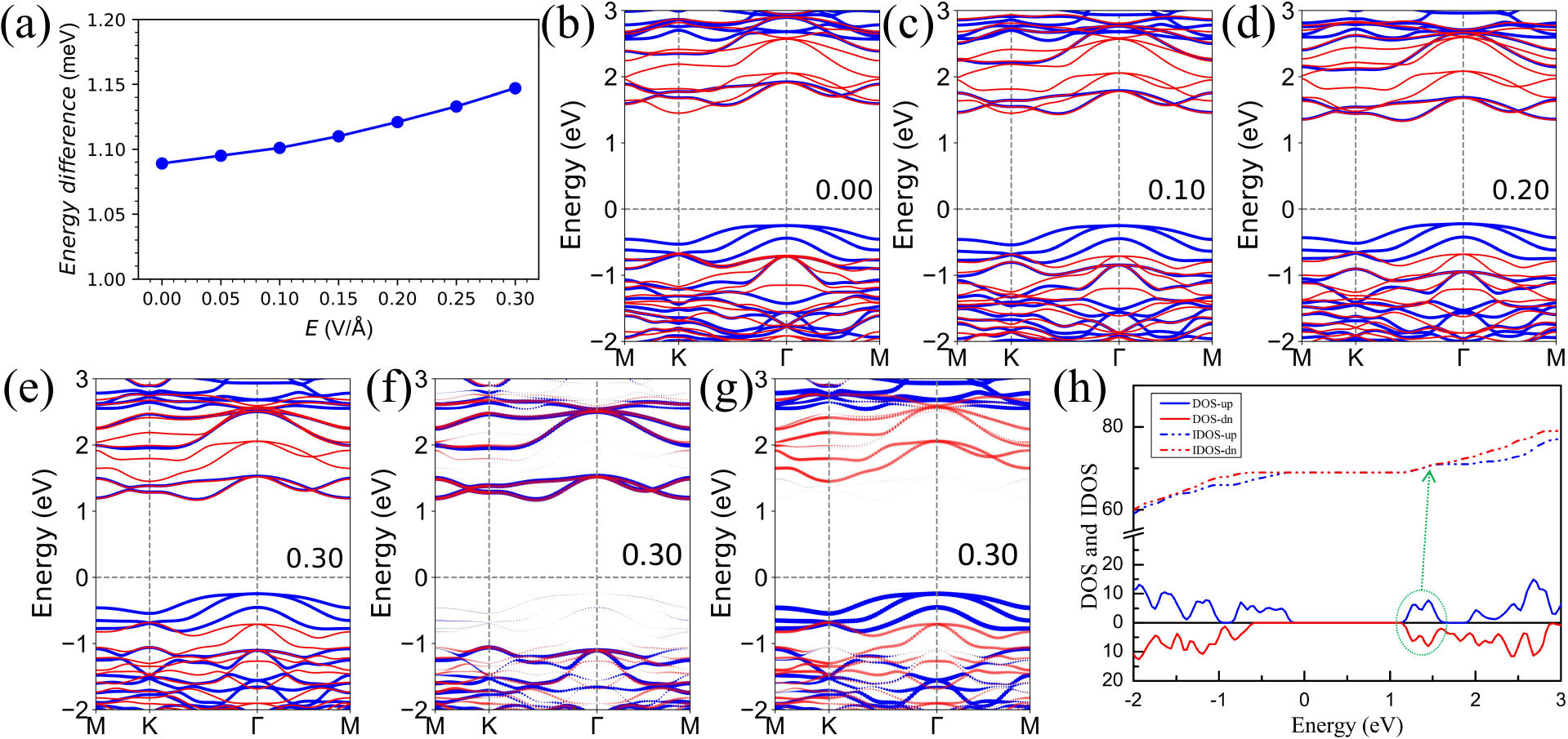}
     \caption{(Color online)For  $\mathrm{Cr_2C_2S_6}$/$\mathrm{CrMoC_2S_6}$ heterojunction, the energy difference  between AFM1 and AFM2 as a function of electric field $E$, with AFM2 taken as the reference (a);  the global spin-polarized band structure at  representative $E$=+0.00 (b), +0.10 (c), +0.20 (d) and +0.30 (e)  $\mathrm{V/{\AA}}$;
 the spin-polarized band structure projected onto $\mathrm{Cr_2C_2S_6}$ (f) and $\mathrm{CrMoC_2S_6}$ (g) at $E$=0.30  $\mathrm{V/{\AA}}$; the spin-polarized DOS and IDOS  at $E$=+0.30  $\mathrm{V/{\AA}}$.  In  (b, c, d, e, f, g, h), the spin-up
and spin-down channels are depicted in blue and red. }\label{c}
\end{figure*}

\textcolor[rgb]{0.00,0.00,1.00}{\textbf{Material realization.---}}We adopt a specific material system to verify the validity of our theoretical model.
Monolayer $\mathrm{Cr_2C_2S_6}$ has been predicted to be a stable $PT$-antiferromagnet\cite{qq3}, which crystallizes in the  $P\bar{3}1m$ space group (No.162).
The monolayer $\mathrm{CrMoC_2S_6}$ can be obtained  via isovalent alloying,  in which one Cr atom in $\mathrm{Cr_2C_2S_6}$
is substituted by Mo atom.  Monolayer $\mathrm{CrMoC_2S_6}$  crystallizes in the  $P312$ space group (No.149),  which  is predicted to be a stable fully compensated ferrimagnet\cite{qq3,f7-1}.   For  $\mathrm{Cr_2C_2S_6}$, two Cr atoms with opposite spins can be connected via $P$ symmetry and $M_{xy}$ symmetry. When one of the Cr atoms is substituted by a Mo atom, the two symmetries disappear in $\mathrm{CrMoC_2S_6}$, giving rise to fully compensated ferrimagnetism.
 The $\mathrm{Cr_2C_2S_6}$ and $\mathrm{CrMoC_2S_6}$   both possesses AFM-N$\acute{e}$el  ground state, and the  optimized  equilibrium lattice constants are  $a$=$b$=5.636 and  5.714 $\mathrm{{\AA}}$\cite{qq3}.  Their band structures  show semiconductor characteristics\cite{qq3}.
Owing to $PT$ symmetry, $\mathrm{Cr_2C_2S_6}$  maintains spin degeneracy across the entire BZ. By contrast, the breaking of
$P$  and $M_{xy}$  symmetries in $\mathrm{CrMoC_2S_6}$  induces FC-FIM spin splitting with $s$-wave symmetry\cite{qq3}.

Next, we investigate the hybrid magnet composed of $PT$-antiferromagnet and fully compensated ferrimagnet.
We directly stack  $\mathrm{Cr_2C_2S_6}$ and $\mathrm{CrMoC_2S_6}$ to construct heterostructures with AA and AB stacking configurations. The calculated results reveal that the AB configuration has lower energy, and its structure is displayed in \autoref{b} (a).
Since the energies of other magnetic configurations of $\mathrm{Cr_2C_2S_6}$ and $\mathrm{CrMoC_2S_6}$ monolayers are much higher than that of the  AFM-N$\acute{e}$el  state ($>$100 meV)\cite{qq3}, and the interlayer interaction of the  heterostructure  is relatively weak (about 1 meV), we only consider two interlayer-modulated magnetic configurations, AFM1 and AFM2 (see \autoref{b} (b, c)). The calculated results indicate that AFM2 is the magnetic ground state, with its energy 1.09 meV lower than that of AFM1.   With AFM2 ordering,  the  optimized  equilibrium lattice constants are  $a$=$b$=5.681 $\mathrm{{\AA}}$.
The global band structure and its projections onto  $\mathrm{Cr_2C_2S_6}$ and $\mathrm{CrMoC_2S_6}$  are plotted in  \autoref{b} (d, e, f).

The calculated results show that the magnetic moments of the two Cr atoms in component  $\mathrm{Cr_2C_2S_6}$ of the heterostructure are 3.00 $\mu_B$ and -3.00 $\mu_B$, respectively, which satisfies the equal-magnitude constraint imposed by
$P$  symmetry\cite{k4}. The magnetic moments of Cr and Mo atoms in component  $\mathrm{CrMoC_2S_6}$ are -2.85 $\mu_B$ and 2.32 $\mu_B$, respectively, consistent with the absence of symmetric coupling between the two magnetic atoms\cite{k6}. Moreover, the total magnetic moment of the heterostructure is strictly 0.00 $\mu_B$, since band gaps exist for both spin-up and spin-down bands throughout the system\cite{k6}.
From the perspective of the global band structure, the heterojunction can be regarded as an FC-FIM semiconductor, since spin splitting exists in the energy bands over the entire BZ and the total magnetic moment of the system is zero. The projected band structure confirms the type-I characteristic of the heterojunction, where the electronic states of conduction and valence bands near the Fermi level are predominantly contributed by component $\mathrm{CrMoC_2S_6}$.
The projected bands of component  $\mathrm{Cr_2C_2S_6}$ indicate that the conduction bands near the Fermi level are nearly spin-degenerate.

To construct a type-II heterojunction, an external electric field can be applied to the $\mathrm{Cr_2C_2S_6}$/$\mathrm{CrMoC_2S_6}$ heterojunction based on our TB model (see \autoref{a1}). We first identify the magnetic ground state under external electric field. \autoref{c} (a) shows that the AFM2 magnetic configuration maintains the ground state in the entire electric field range. The global spin-polarized band structure at  representative $E$=+0.00, +0.10, +0.20 and +0.30  $\mathrm{V/{\AA}}$ along with  the spin-polarized band structure projected onto $\mathrm{Cr_2C_2S_6}$  and $\mathrm{CrMoC_2S_6}$   at $E$=+0.30  $\mathrm{V/{\AA}}$ are plotted in \autoref{c} (b-g).  The calculation results show that,  with the increase of applied electric field, the characteristic conduction band of component $\mathrm{Cr_2C_2S_6}$ shifts toward the Fermi level, while the characteristic valence band of component $\mathrm{CrMoC_2S_6}$ remains almost unchanged, thereby transforming the heterojunction into a type-II alignment.
 According to the projected band structures at $E$=+0.30  $\mathrm{V/{\AA}}$, when the Fermi level shifts appropriately toward the conduction band (electron doping), nearly no net magnetic moment is generated. In contrast, the shift of the Fermi level toward the valence band (hole doping) will induce a net magnetic moment.

To further clarify the regulation effect of different doping types on magnetic moment, the DOS and integrated density of states (IDOS) at
$E$=0.30  $\mathrm{V/{\AA}}$  are presented in \autoref{c} (h).
It can also be seen from the DOS that the spin-up and spin-down states of the conduction band are basically symmetric in the low-energy range, implying spin degeneracy of the band structure over the entire BZ. The total magnetic moment of the system is proportional to the difference between the spin-up and spin-down IDOS. Within the band gap range, the spin-up and spin-down IDOS coincide exactly, indicating that the undoped heterojunction possesses a zero total magnetic moment, which is consistent with the characteristics of FC-FIM semiconductors\cite{k6}. Upon appropriate electron doping, the spin-up and spin-down IDOS remain nearly overlapped, and the total magnetic moment of the heterojunction is still close to zero. By contrast, hole doping leads to an obvious separation between spin-up and spin-down IDOS, which induces a net magnetic moment in the heterojunction. In conclusion, the electric field regulation of the $\mathrm{Cr_2C_2S_6}$/$\mathrm{CrMoC_2S_6}$  heterojunction well validates our theoretical proposal.

\textcolor[rgb]{0.00,0.00,1.00}{\textbf{Discussion and Conclusion.---}}
In fact, we construct zero-net-magnetization hybrid magnets by introducing layer degrees of freedom via heterojunction engineering. Further research is expected to explore intrinsic hybrid magnets, which will facilitate the development and application of this field. The concept of hybrid magnets embodies the idea of hidden physics\cite{h6,h31,h41,h51,h61,h71}, and hybrid magnetic systems can host hidden $PT$-antiferromagnet or hidden altermagnet.
Compared with $PT$-antiferromagnets, altermagnets and fully compensated ferrimagnets exhibit a variety of intriguing physical phenomena, including the anomalous Hall/Nernst effect, nonrelativistic spin-polarized currents, and the magneto-optical Kerr effect\cite{k6,k8}. By introducing the real-space degree of freedom of layer, layer-locked phenomena can be realized, and the related physical properties can be switched on and off via an external electric field.

 In summary, we report the zero-net-magnetization hybrid magnets with oppositely spin-polarized magnetic atoms partially coupled via
$P$,  $C/M$  symmetry or without symmetry. We propose constructing such hybrid magnets using heterojunctions constructed from monolayers of zero-net-magnetization magnets. For type-II aligned heterojunctions, one constituent layer hosts fully compensated ferrimagnet, and  only electron or hole doping can induce a net magnetic moment. Our theoretical proposal is validated by TB simulations on $PT$-antiferromagnet/fully compensated ferrimagnet heterojunctions and first-principles calculations with electric-field modulation on $\mathrm{Cr_2C_2S_6}$/$\mathrm{CrMoC_2S_6}$  heterojunction.
Our work broadens the category of zero-net-magnetization magnets and facilitates the development of zero-net-magnetization spintronics.

\begin{acknowledgments}
This work is supported by Natural Science Basis Research Plan in Shaanxi Province of China  (2021JM-456). We are grateful to Shanxi Supercomputing Center of China, and the calculations were performed on TianHe-2.
\end{acknowledgments}

\end{document}